\documentstyle[11pt,newpasp,twoside,epsf]{article}
\def\edcomment#1{\iffalse\marginpar{\raggedright\sl#1\/}\else\relax\fi}
\reversemarginpar

\begin{document}
\outer\def\ltae {$\buildrel {\lower3pt\hbox{$<$}} \over
{\lower2pt\hbox{$\sim$}} $}
\title{Accretion Flow Along a Dipolar Field: Application to Intermediate Polars}
\author{Jo\~{a}o Batista Garcia Canalle}
\affil{Physics Institute, State University of Rio de Janeiro, \\ 
     Rua S{\~ a}o
Francisco Xavier, 524/3023-D, 
     CEP 20559-900, \\ 
     Rio de Janeiro, RJ, Brazil and \\ 
     Mullard Space Science Laboratory, University College London, \\ 
       Holmbury St Mary, Dorking, Surrey RH5 6NT, United Kingdom } 

\author{Kinwah Wu, Mark Cropper, Gavin Ramsay} 
\affil{Mullard Space Science Laboratory, University College London, \\ 
       Holmbury St Mary, Dorking, Surrey RH5 6NT, United Kingdom} 

\author{Curtis J. Saxton}
\affil{Mount Stromlo and Siding Spring Observatory, \\   
   Research School of Astronomy and Astrophysics, \\
   Australian National University,  ACT 0200, Australia} 

\begin{abstract}

A hydrodynamic formulation for accretion flow channeled by a dipolar
magnetic field is constructed using a curvi-linear coordinate system
natural to the field structure.  We solve the hydrodynamic equations
and determine the velocity, density and temperature profiles of the
post-shock accretion flow.  The results are applied to accretion flows
in intermediate polars.  We have found that for systems with massive
white dwarfs ($\sim 1~{\rm M}_\odot$) the temperature profiles in the
flow can differ significantly to those obtained from models in which
the accretion column is assumed to be cylindrical.

\end{abstract}

\section{Introduction}

In magnetic cataclysmic variables (mCVs), the accretion flow near the
white dwarf is channeled by the magnetic field.  The flow is initially
supersonic but becomes subsonic near the white-dwarf surface, and an
accretion shock is formed.  The shock heats the accreting material to
temperatures of $\sim 10^7 - 10^8$~K.  The shock-heated material is
ionized and emits bremsstrahlung X-rays and cyclotron optical/IR
radiation (Lamb \& Masters 1979).  The height of the shock above the
white-dwarf surface is determined by the cooling processes.  It had
been believed that the height is small ($\sim 10^5-10^6$~cm, see
Wickramasinghe \& Meggitt 1985), and therefore the shock-heated
emission region could be treated as a stack of thin pancake-like
plasmas on the white-dwarf surface.  More recent studies showed that
for some mCV parameters the shock height can be comparable to the
white-dwarf radius when the effects of gravity along the flow are
considered (see e.g.\ Cropper et al.\ 1999).  The latter studies
predicted X-ray spectra in good agreement with the grating spectra of
mCVs obtained by {\it XMM-Newton} and {\it Chandra} (see e.g.\ Cropper
et al.\ 2002).
  
Most existing hydrodynamic calculations of the post-shock accretion
flow in mCVs assume a simple geometry, such as a semi-infinite slab or
a cylinder (Aizu 1973; Chavelier \& Imamura 1982; Wu, Chanmugam \&
Shaviv 1994; Cropper, Ramsay \& Wu 1998).  These simple geometries are
good approximations, provided that the shock height is small and the
flow is azimuthal symmetric and aligned close to the polar axis. For
accretion flows in intermediate polars, these two conditions are not
always satisfied, and hence, the effect of the magnetic-field geometry
needs to be investigated in order to obtain accurate velocity and
temperature profiles of the accretion flow.

Here, we present a hydrodynamic formulation in terms of a curvi-linear
coordinate system natural to channeled flows in dipole magnetic
fields.  We solve the hydrodynamic equations and determine the
velocity, density and temperature structures of the post-shock region.
The results are compared with those of Cropper et al.\ (1999), in
which a azimuthal cylindrical accretion column is assumed.
 
\section{Formulation}
 
\begin{figure}  
\plottwo{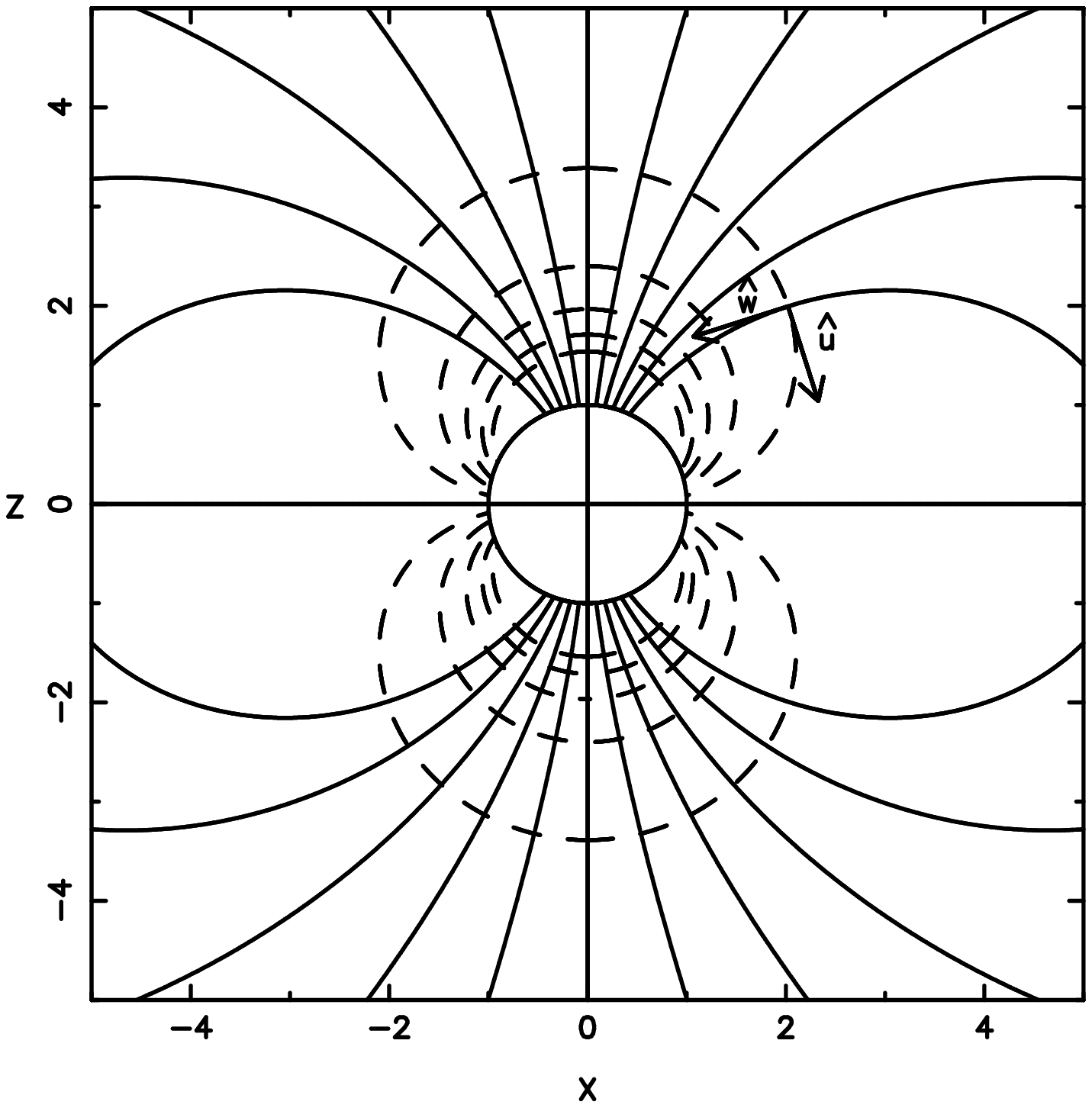}{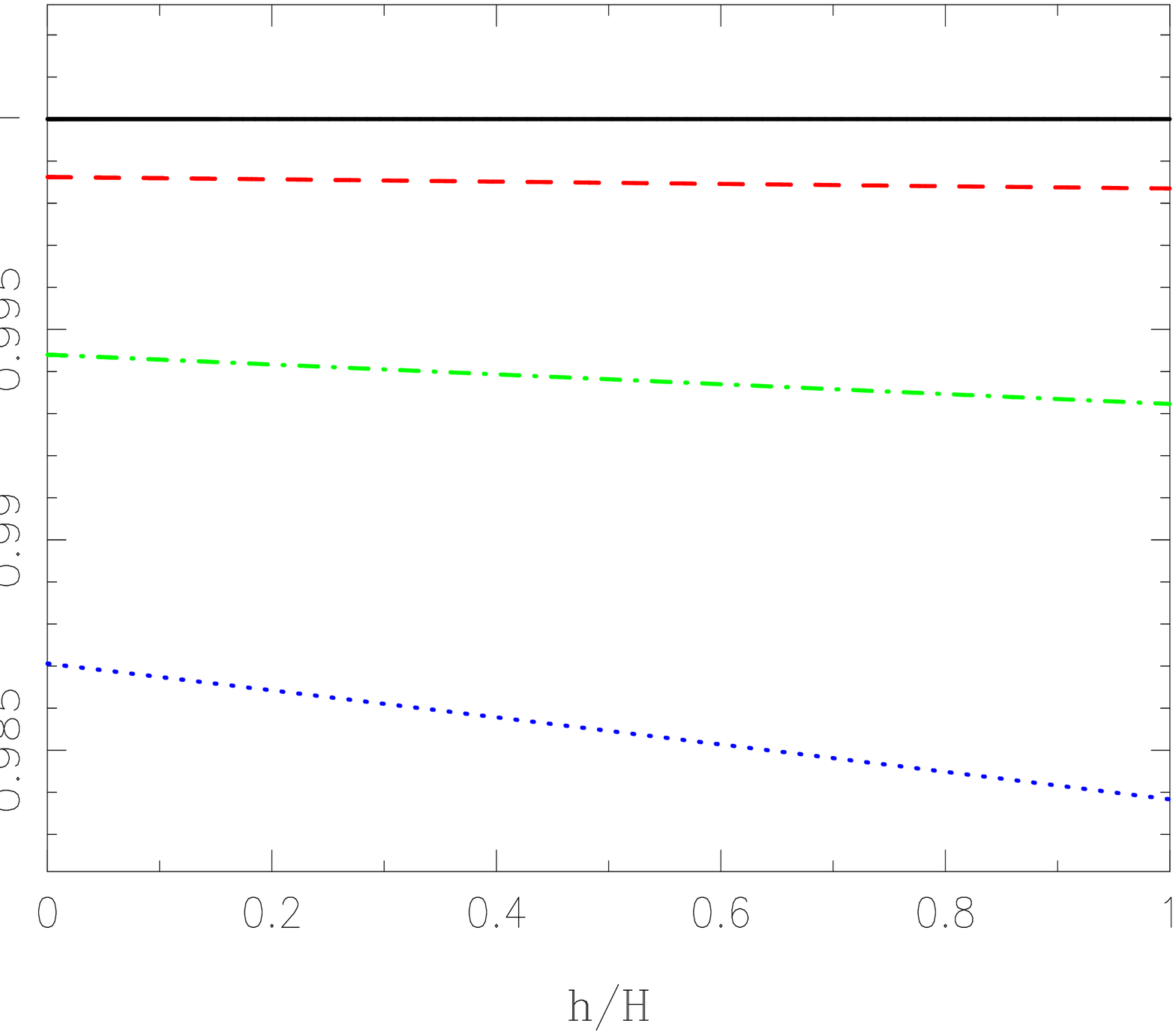} 
\caption{(Left) The ($u,w$) plane of the curvi-linear coordinate
system used in the formulation.  Solid curves represent dipole
magnetic field lines, and dashed curves represent field equi-potential
surfaces.  (Right) The $w$ component of the gravity force, $g_{w}$, as
a function of height $h$, for field lines with different foot points.
The force component is normalized to that of the field line along the
polar axis, $g$, and the height is normalized to the shock height $H$.
The lines (from top to bottom) correspond to the gravity force along
the field lines with foot points at magnetic co-latitude $\theta =
0^{\circ}$, $6^{\circ}$, $12^{\circ}$ and $18^{\circ}$ respectively.}
\end{figure}    

We assume the accreting material is an ideal gas, with a pressure $P$,
density $\rho$ and temperature $T$, related by $P = {\rho\,k_{\rm B}
T}/{\mu\, m_{\rm H}}$, where $k_{\rm B}$ is the Boltzmann constant,
$m_{\rm H}$ is the hydrogen mass, and $\mu$ is the mean molecular
weight. In stationary accretion, the hydrodynamic equations governing
the flow are

\begin{eqnarray}  
\nabla \cdot (\rho \mbox{\bf v}) &  =  & 0 \ ,  \nonumber \\ 
(\mbox{\bf v}\cdot \nabla) \mbox{\bf v} + \frac{1}{\rho}\nabla P 
  & = &  \mbox{\bf g} \ , \nonumber \\ 
(\mbox{\bf v}\cdot \nabla) P 
   - \frac{\gamma P}{\rho}(\mbox{\bf v}\cdot \nabla )\rho 
  & =  &-(\gamma -1)\Lambda  \ , \nonumber 
\end{eqnarray}    

\noindent where ${\bf v}$ is the velocity of the flow, $\gamma$ is the
adiabatic index of the gas, $\mbox{\bf g}$ is the gravitational
acceleration, and $\Lambda$ is the effective cooling function.  (See
Wu (2000) and references therein for the hydrodynamic formulation of
accretion flow in mCVs.)
  
We consider a coordinate system ($u$, $w$, $\varphi$), in which the
orthogonal unit vectors $\hat u$, $\hat w$ and $\hat \varphi$ are
defined as follows: $\hat u$ is on the equi-potential surface of the
field at a fixed azimuthal angle, $\hat w$ is along the magnetic field
line, and $\hat \varphi$ is the same as that for the azimuthal
coordinate in the spherical coordinate system (see Fig.~1, left
panel).  The metrics of the coordinate system, $h_{1}$, $h_{2}$ and
$h_{3}$, are given by

\begin{eqnarray}
  h_{1} & = &  \sqrt{   \left(\frac{\partial x}{\partial u}\right)^{2}
         +\left(\frac{\partial y}{\partial u} \right)^{2}
         +\left(\frac{\partial z}{\partial u}\right)^{2} }   \ , \nonumber \\ 
  h_{2} & = & \sqrt{   \left(\frac{\partial x}{\partial w}\right)^{2}
         +\left(\frac{\partial y}{\partial w} \right)^{2}
         +\left(\frac{\partial z}{\partial w}\right)^{2} }   \ ,  \nonumber \\ 
  h_{3} & = & \sqrt{   \left(\frac{\partial x}{\partial \varphi}\right)^{2}
         +\left(\frac{\partial y}{\partial \varphi} \right)^{2}
         +\left(\frac{\partial z}{\partial \varphi}\right)^{2} }  \ ,  \nonumber  
\end{eqnarray}   

\noindent where ($x,y,z$) is the Cartesian coordinate system.  We may then
express the hydrodynamic equations in terms of ($u,w,\varphi$).

We integrate the mass continuity equation directly and obtain $h_1 h_3
\rho v = C$, where $C$ is a constant.  Substituting this into the
other two hydrodynamic equations yields

\begin{eqnarray}
   \frac{\partial \xi}{\partial w}  
  &  = & -h_{2}\left[\frac{g_{w}}{v} -
    \frac{\cal H}{h_{2}}(\xi -v) \right] \ , \nonumber  \\  
  \frac{\partial v}{\partial w}  & 
  = & - \left(\frac{h_{2}}{\gamma \, (\xi - v) - v} \right) 
  ~\left[{\frac{\gamma -1}{C}\,
    h_{1}h_{3}\Lambda \,  +
   \frac{\gamma {\cal H}}{h_{2}} (\xi - v)v} - g_{w} \right]  \ .  \nonumber 
\end{eqnarray}  

\noindent Here, $\xi \equiv v + ({P}/{\rho v})$, $g_{w}$ is the
gravitational acceleration projected on $\hat w$ (Fig.~1, right
panel), and the function ${\cal H}(u,w) = {\partial}\ln(h_1
h_3)/{\partial w}$, describes the change of cross-section area of the
flow-flux tube.

Our cooling function $\Lambda$ consists of two parts, one for
bremsstrahlung cooling and another for cyclotron cooling.  We adopt
the composite cooling function of Wu (1994) and Wu et al.\ (1994)
(see also Cropper et al.\ 1998).  As cyclotron cooling is practically
unimportant in intermediate polars, we simply set the efficiency
parameter $\epsilon_{\rm s} =0$, i.e.\ $\Lambda = A \rho^2
(P/\rho)^{1/2}$, where the constant $A = 3.9\times 10^{16}$ in
c.g.s. units.  We assume a cold stationary wall for the lower boundary
condition as in Chevalier \& Imamura (1982) and Wu et al.\ (1994).
For the upper boundary condition a strong adiabatic shock is
considered, but with this coordinate system its location must be
determined self-consistently.  Detailed treatment of the upper
boundary condition will be presented in Canalle et al.\ (2003).  With
the cooling function and the boundary conditions defined, we can now
integrate the hydrodynamic equations along the field lines numerically
and obtain the velocity, density and temperature profiles of the flow.
   
\section{Results and Discussion}

\begin{figure}   
\plotone{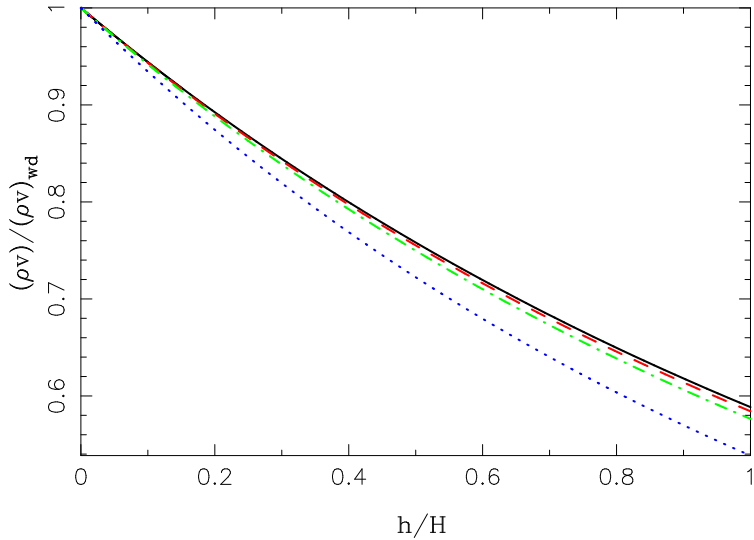}  
\caption{The quantity $\rho v$ as a function height along the field
lines, normalized to the values, $(\rho v)_{\rm wd}$, at the
white-dwarf surface. From top to bottom, the foot points of the field
lines are at magnetic co-latitudes $\theta = 0^{\circ}$, $6^{\circ}$,
$12^{\circ}$ and $18^{\circ}$ respectively.}
\end{figure}    

In all calculations, unless otherwise stated, the white-dwarf mass
$M_{\rm wd} = 1{\rm M}_\odot$ and the specific mass accretion rate
$\dot m = 2.0 \hspace{1mm}{\rm g~cm}^{-2}{\rm s}^{-1}$.  The
white-dwarf radius, $R_{\rm wd}$, is determined by the Nauenberg (1972)
mass-radius relation.

For comparison we also calculate the velocity, temperature and density
profiles of a flow with an azimuthal symmetric cylindrical accretion
column using the formulation in Cropper et al.\ (1999).  The two main
differences in the two formulations are that for flows channeled
strictly by dipolar magnetic fields the cross section of the flow-flux
tube decreases with distance to the white-dwarf surface and that the
gravitational acceleration varies in strength and direction along the
flow.  As shown in Figure 1 (right panel), for the parameters assumed
the gravitational accelerations are smaller for the flows along field
lines with foot point co-latitudes $\theta$ farther away from the
pole.  With the parameters assumed in Figure 1, the value of
$g_w/|{\bf g}|$ at the shock is about 0.985 for $\theta = 18^{\circ}$:
the value can be much smaller for larger $\theta$ and also for
parameters which yield shock heights significantly above the
white-dwarf surface.  Figure 2 shows the $\rho v$ profiles for various
$\theta$.  For small $\theta$, the profiles are similar, but when
$\theta$ is substantially large the deviation becomes significant:
compare the profiles of $\theta = 12^{\circ}$ and $18^{\circ}$ with
those of smaller $\theta$.

\begin{figure}   
\plotone{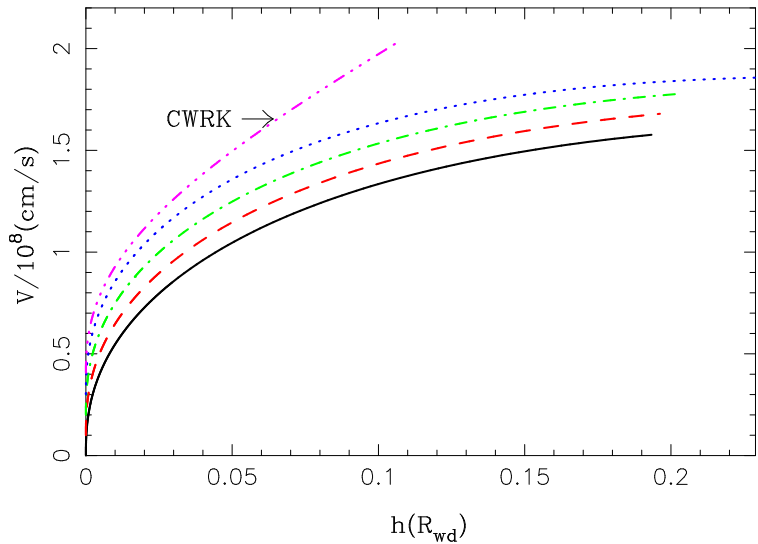}  
\caption{The velocity profiles of the post-shock accretion flow.  The
solid, dashed, dash-dotted and dotted curves correspond to flows along
field lines with foot points at magnetic co-latitudes $\theta =
0^{\circ}$, $6^{\circ}$, $12^{\circ}$ and $18^{\circ}$ respectively
The dash-dot-dot-dotted curves curve (labeled CWRK) correspond to the
velocity profile obtained from the calculation following Cropper et
al.\ (1999).  The curves are terminated at the heights where the
shocks are located.  The velocities of the curves are shifted
progressively upward each by $0.1 \times 10^8$~cm~s$^{-1}$ to amplify
their differences.}
\end{figure}   

Figure 3 shows the velocity profiles of the flows.  The shape of the
profiles for the flows are quite similar regardless of the differences
between the values of $\theta$.  However, the shock heights are
different, and hence the shock temperatures are different accordingly.
Interestingly, the profiles are quite different to those of the
cylindrical accretion column (Cropper et al.\ 1999).  In particular,
the latter yield a smaller shock height, thus a hotter and more
compact post-shock emission region.  This will affect the properties
of the optical/IR cyclotron radiation as they are emitted mainly from
the hotter part of post-shock region.  The effects on the
bremsstrahlung X-ray continuum may be less affected as it is emitted
from the cooler bottom of the post-shock region.
    
Figure 4 shows the temperature profiles for systems with various
white-dwarf masses.  Both the flows in cylindrical columns and flows
confined by dipolar fields show that the shock temperatures increases
with the white-dwarf mass. For white-dwarf with masse \ltae
0.7~M$_\odot$, the profiles for the two cases are similar, with the
model with cylindrical columns giving slightly higher
temperatures. However, when the white-dwarf mass is sufficiently large
($\sim 1.0$~M$_\odot$), the differences between the two cases becomes
obvious.  Also, the model with cylindrical columns tends to over
predict the gas temperatures. The difference between the two cases can
be explained qualitatively as follows. Larger white-dwarf masses
result in higher shock temperatures and hence larger shock heights
above the white-dwarf surface.  When the shock height becomes too
large, the cylindrical accretion column is not a very good
approximation, and the variation in the gravitational acceleration
along the flow becomes significant. The field geometry has stronger
effects on the accretion onto massive white dwarfs than low-mass white
dwarfs. Therefore, when modelling the X-rays and optical/IR radiations
from the shock-heated accreting gas in the systems with massive white
dwarfs, one must consider the effects due to the field geometry.

\begin{figure}   
\plotone{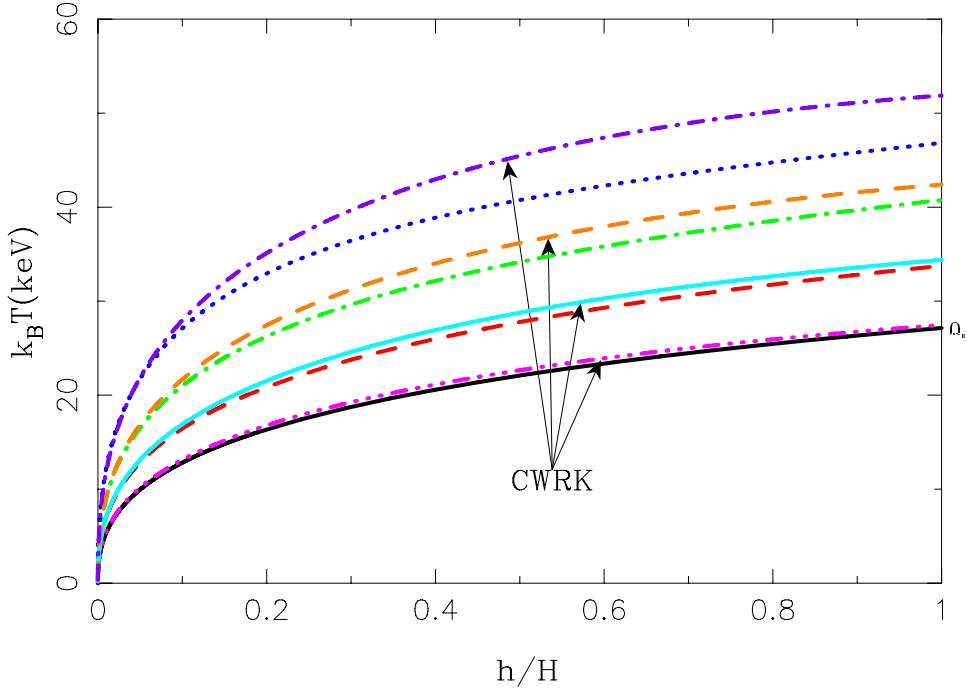}  
\caption{Temperature profiles of flows along a dipolar field in
comparison with flows assuming a cylindrical accretion column (labeled
CWRK, Cropper et al.\ 1999) for white-dwarf masses 1.0, 0.9, 0.8 and
0.7~M$_\odot$ (curve pairs from top to bottom). For the flows along a
dipolar field, the foot point of the field line is fixed at the
magnetic co-latitude $\theta = 18^{\circ}$. In all cases, the mass
accretion rate $\dot m = 2.0~{\rm g~cm}^{-2}{\rm s}^{-1}$.}
\end{figure}

These finding may explain the finding of Cropper et al (1999) and
Ramsay (2000) who found that the mass of the white dwarf in mCVs were
biased towards higher masses compared to isolated white dwarfs when
they fitted X-ray spectra with the cylindrical formulation. Work is
currently in progress to implement the affect of the field geometry
and determine how it will affect the resulting white dwarf masses.

In summary, we present a hydrodynamic formulation using a curvi-linear
coordinate system defined by the dipole field configuration.  We use
this formulation to calculate the velocity, density and temperature
profiles of the accretion flows in mCVs.  We have shown that effects
of field geometry are important for mCVs with high-mass white dwarfs.
Our formulation improves the calculations of temperature and density
structures of post-shock emission region in mCVs, in particular, the
intermediate polars, and thereby predict more accurately the X-ray and
optical/IR emission from these systems.

\vspace{1cm}
{\bf \noindent Acknowledgments}

\noindent JBGC acknowledges the financial support from the Conselho Nacional de
Desenvolvimento Cientifico e Tecnologico (CNPq) and to the State
University of R io de Janeiro (UERJ) for the leave of one year inside
the program PROCAD to take part in this research.

\end{document}